\documentclass[fleqn,twoside]{article}
\usepackage{espcrc2}

\usepackage{graphicx}

\title{Precision muon lifetime and capture experiments at PSI}

\author{F.E.~Gray\address[Berkeley]{Department of Physics, University of California, Berkeley; Berkeley, CA~~94720, U.S.A.}
\thanks{This work is supported in part 
by the U.S. Department of Energy and the National Science Foundation.
The author gratefully acknowledges travel funds from the
NuFact04 Young Scientist Support Program.}\\
on behalf of the $\mu$Lan\thanks{
K.M.~Crowe, F.E.~Gray, and B.~Lauss (U. of California, Berkeley and LBNL);
R.M.~Carey, W.~Earle, A.~Gafarov, B.~Graf, M.~Hance, M.~Hare
% ...
(Petersburg Nuclear Physics Inst.).
} Collaborations.
}

\begin{document}
\begin{abstract}
The $\mu$Lan experiment at the Paul Scherrer Institute will measure the 
lifetime of the positive muon with a precision of 1~ppm, giving a value 
for the Fermi coupling 
constant $G_F$ at the level of 0.5~ppm.  Meanwhile, by measuring the observed 
lifetime of the negative muon in pure hydrogen, the $\mu$Cap experiment will
determine the rate of muon capture, giving the proton's pseudoscalar  
coupling $g_p$ to 7\%.  This coupling can be calculated precisely from heavy 
baryon chiral perturbation theory and therefore permits a test of QCD's 
chiral symmetry.
\end{abstract}
\maketitle

\section{Muon lifetime}

The muon lifetime $\tau_\mu$ is closely related to the Fermi coupling
constant $G_F$, which sets the strength of the weak interaction:
\begin{displaymath}
\frac{1}{\tau_{\mu}} = \frac{G_F^2 m_\mu^5}{192 \pi^3} (1 + \delta q) ~.
\end{displaymath}
The term $\delta q$ includes the QED radiative corrections which, until
1998, were only known to a precision of 30 ppm.  Van Ritbergen and 
Stuart~\cite{vanRitbergen:1999fi} have now calculated these corrections
through two-loop order, with a residual uncertainty at the level of 0.3~ppm.
Consequently, the determination of $G_F$ is now limited only by our knowledge
of the muon lifetime at the 18~ppm level.

There is a deep connection between $G_F$ and the Higgs vacuum expectation 
value, represented by the relation~\cite{okada}
$G_F = 1/\sqrt{2} v^2$, which may be read as ``the weak interaction is the 
generator of mass in the universe.''  It therefore becomes clear 
that $G_F$ is a truly fundamental parameter of the standard 
model.  To take an example of its impact, the LEP electroweak working 
group~\cite{lepewwg} performs a global fit to the standard model using $G_F$ 
together with the LEP observables such as $M_Z$ and $\alpha$ and is able to 
determine the top quark mass with a precision of 9.7~GeV~(5.4\%) without 
producing a single real top quark.  As a fundamental constant, $G_F$ should be 
measured as precisely as possible with today's technology, even though a 
corresponding improvement in the precision of $M_Z$ will probably await the 
construction of a linear collider or muon collider.  Our collaboration also 
has a more immediate motive for measuring the muon lifetime: we will use 
the $\mu^-$ capture rate in hydrogen and deuterium as a probe of the internal 
structure of the proton and the deuteron, as we discuss in the second part 
of this paper.  A precise measurement of the $\mu^+$ lifetime will directly 
improve the precision of these capture rates.

The standard technique to measure the muon lifetime is to stop low-energy 
muons in a target, observe the Michel electrons from muon decay, and
fit the exponential distribution of $t_e - t_\mu$, the time between the
muon stop and its decay.  Previous experiments have operated at muon 
rates of order $1/(10\tau_\mu) \simeq $25~kHz, where typically only a single 
muon is present in the target at a time; an offline cut suppresses
events where two muons arrive too close together.  However, it would be 
difficult to make a ppm-scale measurement in this mode, since it would take 
more than 250 days of operation to record the requisite $10^{12}$ events.  
A number of efforts have attempted to solve this problem by ``parallel 
processing'' of muons.  Like the RIKEN-RAL experiment~\cite{tomono}, 
$\mu$Lan uses a pulsed beam structure, stopping an ensemble of muons in the 
target and observing their decays.  The muon bunches are, however, much 
smaller in the case of $\mu$Lan, typically 20~muons at a time, rather than 
of order $10^4$ as at RIKEN-RAL, so the systematic uncertainty due to 
overlapping pulses should be dramatically smaller.
The time structure, optimized for measuring the muon lifetime, is obtained 
using a fast kicker to modulate the continuous beam available at PSI.
This approach is in contrast with that of the FAST experiment~\cite{fast}, 
also conducted at PSI, which stops pions continuously in a finely segmented 
scintillating fiber target, separating muons by space rather than by time.  

The kicker will cycle between beam-on and beam-off states, accumulating
muons in the target for 5~$\mu$s and observing their decays for the 
next 22~$\mu$s.  The beam-off state is achieved by charging two pairs of 
plates, each $(75 \times 20)$~cm$^2$, to a $\pm12.5$~kV potential, deflecting 
the beam into an absorber.  The rise and fall times are expected to be 45~ns, 
limiting the time in which muons are deflected to a poorly-defined position.
Tests with a static field on the kicker plates suggest that an extinction 
factor of greater than 300 will be achieved.  The beam will be continuously 
monitored by a wire chamber that has been optimized for high rates to ensure 
the stability of this extinction factor on both long and short time scales.
The beam is collimated upstream of the kicker to a rate of 12~MHz, 
giving 2~MHz of muons at the target during the beam-on period.

\begin{figure}
\begin{center}
\includegraphics[width=0.45\textwidth]{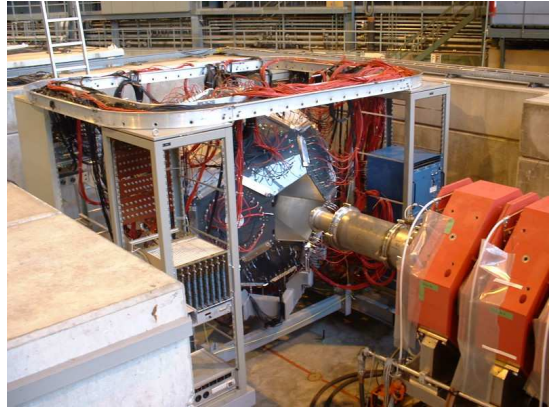}
\vspace{-1cm}
\end{center}
\label{mulan-photo}
\caption{The $\mu$Lan detector in place in the $\pi$E3 beamline at PSI.}
\end{figure}

The $\mu$Lan detector (Figure~\ref{mulan-photo}) is a truncated 
icosahedron (``soccer ball'') surrounding the stopping target, with two 
pentagons removed to permit the beam to enter and exit.  It consists of 
170 nested pairs of triangular plastic scintillator tiles.
A coincident signal in the inner and outer tiles 
of a pair will be required in order to reduce accidental backgrounds.
These detectors give on average 80 photoelectrons per minimum ionizing
particle in each layer.

The muon's spin precesses in any external magnetic field. 
Because parity violation in muon decay sends the electrons
preferentially along the axis defined by the spin, any
geometric inhomogeneity in the electron detector acceptance distorts the
measured time spectrum.  We reduce this effect in several ways.
The detector covers a large solid angle and
is as spherically symmetric as practical to permit cancellation of the
asymmetry in the sum of each opposing pair.  For most of our running time, we
will use a target made of a ferromagnetic alloy (Arnokrome-3) with
large ($\sim$4~kG), nonuniform internal magnetic fields.  As a control, we can
swap it with a silver target disk, which preserves
the polarization.  Finally, we apply a uniform external magnetic field of
approximately 120~G; each of the muons within a bunch arrives at a different
time, so their spin motion in this field is dephased.

The high segmentation of the detector reduces the impact of overlapping
pulses (``pileup'').  It will be further reduced by recording
each detector's output with a waveform digitizer (WFD), 
using a flash ADC to sample the signal at 500~MHz.  The resulting waveforms
will be reduced to times and amplitudes by an online computing ``farm,''
which should be able to resolve pulses at separations as close as 
$\delta t \sim 4$~ns based on a fit to the pulse shape.  The separation 
in pulse amplitude between single and double hits will provide an additional
pileup suppression factor of $\sim$25.  The final systematic
uncertainty from overlapping pulses should be less than 0.1~ppm.

Following a number of successful beam and detector test runs, $\mu$Lan will
have its first data collection period this fall with a kicked beam.  Because 
the WFD construction has been delayed, we will have to use 
conventional discriminators and multi-hit TDCs, with the goal of reaching 
a 3~ppm measurement of $\tau_\mu$.  In 2005, we will have a major production 
run with the WFD to reach the full proposed $10^{12}$ statistics.

\section{Muon capture}

The current associated with the weak interaction between a muon and a
``free'' quark inside the proton is a maximally parity-violating $V-A$ form, 
proportional to $\gamma_\mu (1-\gamma_5)$.  However, the QCD binding of quarks 
within the proton adds several induced form factors to the problem; the 
most general form of the interaction between a muon and a proton 
becomes~\cite{Gorringe:2002xx}
\begin{eqnarray*}
   J_\alpha &=&  g_v(q^2) \gamma_\alpha + i \frac{g_m(q^2)}{2 m_N} \sigma_{\alpha\beta} q^\beta \\
   & &   - g_a(q^2) \gamma_\alpha \gamma_5 - g_p(q^2) \frac{q_\alpha}{m_\mu} \gamma_5 \\
   & &   + {\textrm {second class currents}} ~.
\end{eqnarray*}
The first three form factors ($g_v$, $g_a$, and $g_m$) are known very precisely,
to at worst a few tenths of a percent, but there is considerable 
experimental ambiguity in the case of $g_p$, which is only known to about 20\%.

However, there are reliable theoretical predictions.
A heavy baryon chiral perturbation theory calculation 
gives~\cite{chipt}
\begin{displaymath}
 g_p(q^2) = \frac{2 m_\mu g_{\pi NN} F_\pi}{m_\pi^2 - q^2} - \frac{1}{3} g_a(0) m_\mu m_N r_A^2 
\end{displaymath}
\begin{displaymath}
 g_p(q^2 \rightarrow (p_n - p_p)^2)  = 8.26 \pm 0.23 ~(\pm3\%) ~.
\end{displaymath}
This calculation, an expansion parameterized by the light quark mass,
is a precise result of a low-energy expression of the fundamental chiral 
symmetry of QCD, so it is extremely important to verify it.
Furthermore, it is consistent with older partially conserved axial 
current (PCAC) calculations, which give
\begin{displaymath}
 g_p(q^2) = \frac{2 m_\mu m_N}{m_\pi^2 q^2} g_a(0) \\
\end{displaymath}
\begin{displaymath}
 g_p(q^2 \rightarrow (p_n - p_p)^2)  = 8.7 ~.
\end{displaymath}

Typically, the negative muon decays 
just as the positive muon does.  
However, in hydrogen gas, it has an additional possibility:
about 0.15\% of the time, it is instead captured by the proton.
It is this 0.15\% value that we aim to measure to high precision by 
comparing the lifetimes of positive and negative muons in hydrogen
and computing $\Lambda_s = 1/\tau_{\mu^-} - 1/\tau_{\mu^+}$.  Each of the 
lifetimes will be measured to 10~ppm by $\mu$Cap; a significant improvement in
precision could be obtained by instead taking $\mu$Lan's 1~ppm
measurement of $\tau_{\mu^+}$.
The quantity $\Lambda_s$ may in turn be related to $g_p$~\cite{commins}; 
a 1\% measurement of $\Lambda_s$ should give $g_p$ to 7\%.  

Previous measurements of the muon capture process have been 
limited by the molecular physics of the liquid hydrogen target.   
Specifically, the transition rate $\lambda_{op}$ between the ortho- and para- 
states of the muonic hydrogen molecule is not well-determined, 
and the capture rates from these two states are quite different.
$\mu$Cap is much less sensitive to these effects because it uses a 10~bar
hydrogen gas target, which has only 1\% of the density of liquid hydrogen.
The muons initially statistically populate the atomic hyperfine singlet and 
triplet states, but quickly drop through thermal collisions to the singlet 
state and stay there rather than entering molecular states.

The hydrogen is contained in a time projection chamber (TPC) where 
it is used as both the target material and the active chamber gas.
In the TPC, each muon is tracked to its stopping point where the Bragg
peak is identified, guaranteeing that it stopped in hydrogen.  The TPC is
surrounded by a cylindrical electron detector with two wire chambers and
two layers of scintillator.

The hydrogen gas must be very pure, since muons would transfer to and
subsequently capture on heavier impurity nuclei, and the capture rate is
roughly proportional to $Z^4$.
The TPC is constructed of materials that can be baked in vacuum 
to 130$^\circ$~C to reduce outgassing.  The protium is filled through a 
palladium filter, and a system has been constructed to continuously circulate 
it through a Zeolite absorber cooled by liquid nitrogen.  
An in-situ analysis is performed by searching for a second track in the TPC 
corresponding to the recoil nucleus following a muon capture on the impurity.
This technique supplements an external chromatographic analysis of the gas.

Deuterium is more problematic than higher-$Z$ impurities.  Typically, a $\mu$d 
atom diffuses a long distance from the muon stopping point; there is a 
minimum of the $\mu$d-p scattering cross section 
at 1.6~eV, giving it a long mean free path.  It therefore leaves the fiducial 
volume and is probably captured in the TPC frame.  The recoil nucleus is not 
observed; however, it may be possible to make an in-situ measurement by 
tracking the electrons back to the decay vertex and relating the concentration 
to the distribution of drifts from the muon stop to decay.  Also, if the TPC 
gain can be increased sufficiently, it may be possible to identify muon 
catalyzed fusion events where
$\mu d + p \rightarrow \mu$~(5.3 MeV)~$+ ^3$He~(0.2 keV).

\begin{figure}
\begin{center}
\includegraphics[width=0.45\textwidth]{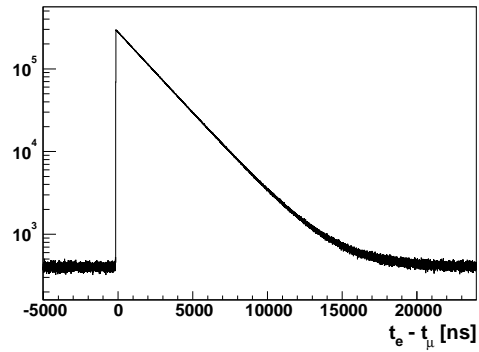}
\vspace{-1cm}
\end{center}
\label{mucap_time_spectrum}
\caption{Time spectrum from the 2003 $\mu$Cap run including $5\times10^8$ $\mu^-$ decays.}
\end{figure}

In 2003, we collected $5 \times 10^{8}$ clean muon decay 
events (Figure~\ref{mucap_time_spectrum}), which 
would lead to a statistical uncertainty of about 5\% on the capture rate.
The systematics, especially from deuterium, are still being evaluated. 
This autumn, we intend to collect 
an order of magnitude more data, amounting to half of the total proposed 
statistics of $10^{10}$ muons.  The upgrades relative to last year's run
include the gas circulation system, the outer wire chamber in the electron
detector, and somewhat higher gain in the TPC.  We will then complete the 
proposed data-taking in 2005.  In 2006 and beyond, we will consider
operating the experiment in a ``muon-on-request'' mode with the $\mu$Lan
kicker to increase the rate at which statistics can be collected. 
We will also continue to investigate the feasibility of modifying our 
apparatus to measure the muon capture rate in deuterium, which would 
provide a sensitive test of two-body currents and constrain the effective
field theory parameter $L_{1,A}$, an important ingredient in the
absolute neutrino flux at SNO~\cite{Chen:2002pv}.

\end{document}